\documentstyle[11pt]{article}
\newcommand{\be}{\begin{equation}}
\newcommand{\ee}{\end{equation}}

\newcommand{\la}[1]{\label{#1}}
\newcommand{\r}[1]{(\ref{#1})}

\begin{document}

\title{Meson bound states in multiflavour massive
Schwinger model\thanks{
Work supported by the KBN
grants no.\ 2P 30207607 and 2P 30204905.}}
\author{
{\sc M.\,Sadzikowski}\thanks{Fellow of the Polish Foundation (FNP)
scholarship for the year 1996}\\
{\it Institute of Nuclear Physics,} \\
{\it Radzikowskiego 152, PL-31\,342 Krak\'ow, Poland}\\
and\\
{\sc P.\,W\c{e}grzyn}\thanks{E-mail {\sc wegrzyn@ztc386a.if.uj.edu.pl}}\\
{\it Institute of Physics, Jagellonian University,} \\
{\it ul.\,Reymonta 4, PL-30\,059 Krak\'ow, Poland}}
\date{}

\maketitle
\begin{center}

\begin{abstract}

The problem of meson bound states with $N_f$ massive fermions
in two dimensional quantum electrodynamics is discussed.
We speculate about the spectrum of the lightest particles by means of
the effective semiclassical description. In particular, the systems
of fundamental fermions with $SU(2)$ and $SU(3)$ flavour symmetries
broken by massive terms are investigated.
\end{abstract}
\end{center}
\newpage

\section{Introduction}

The literature on the Schwinger model \cite{schwinger} and its generalizations
should refer to at least a hundred of interesting and important papers.
This theoretical model is usually used to demonstrate some important
phenomena known from more realistic models. It is easy to see and
investigate here screening and confinement, 
$U(1)$ problem, chiral symmetry breaking
and topological vacua,  formation of bound 
states and
finite temperature effects. On the other hand,
known non-perturbative techniques tested here,  like
lattice calculations or sum rules,  
reveal many pitfalls, what makes their use doubtful.
 Surprisingly, the two-dimensional
Schwinger model shares many phenomenological features in common with 
four-dimensional QCD. While one can state that the problem of massless 
charged fermions in two dimensions has been elaborated in all details 
\cite{lovenstein}--\cite{seiler}, the
situation when fermion masses are finite \cite{coleman2}--\cite{gross}
still requires further investigations.
In particular, it would be interesting to understand in the theory
with the confinement the mechanism of formation of the lightest physical
particles from the fundamental fermions. This subject is discussed in
this paper. 

Our paper is organized as follows: in the Section 2, we introduce
briefly the multiflavour massive Schwinger model in
its bosonized version. Classical equations of motion
and ground states are described. In the Section 3, we discuss the
lightest meson bound states in the case when fundamental
fermions ('quarks') are heavy (or in other words the coupling constant 
is weak). To make our paper self-contained, in the first part of
this section we remind the most important points of the semiclassical
quantization procedure applied to the 'particle-like' solutions of
the sine--Gordon theory. This knowledge is used in subsequent
developments. In the  Section 4, the analysis of the lightest
meson states is extended to the case of light quarks (the strong
coupling regime). We discuss the model for $SU(2)$ and $SU(3)$ flavour
groups separately, both in unbroken and broken cases. 
 In the Appendix,
we advertise the method useful to derive approximately particle-like
solutions to the classical field equations. This method is explained
on the simplest example of the sine--Gordon theory, but its
advantage is that it can be used to more involved problems.

\section{Lagrangian}

We will consider $N_f$ fundamental charged and massive Dirac fermions in 
(1+1)-dimensional
Minkowski spacetime, with the following Lagrangian density
\be
\la{fermla}
{\cal L} = -\frac{1}{4} F_{\mu\nu }F^{\mu\nu} + \sum_{a=1}^{N_f}\bar{\psi^a}(
i\gamma^\mu\partial_\mu - e\gamma^\mu A_\mu - m_a)\psi^a ,
\ee
where 
\be
F_{\mu\nu } = \partial_\mu A_\nu - \partial_\nu A_\mu ,
\ee
$e$ is the coupling constant (dimension of mass). 
We allow for masses of fermions
$m_a$ to be different for each flavour $a$. In the exactly solvable
massless case $m_a=0$, the classical system possesses a global symmetry
group $U(N)_L\times U(N)_R = U(1)_V\times U(1)_A\times SU(N)_L\times SU(N)_R$.
At the quantum level, the axial symmetry $U(1)_A$ is broken down by the
anomaly. In the massive case, the flavour symmetry $SU(N)_L\times SU(N)_R$
is broken explicitly to $U(1)^{N_f-1}$, leaving only $(N_f-1)$ conserved
flavour numbers. We are going to make use of 
the standard (Abelian) bosonization rules,
\be
N_{m_a}[\bar{\psi^a}\gamma^\mu\psi^a]=\frac{1}{\sqrt{\pi }}\epsilon^{\mu\nu }
\partial_\nu\Phi^a ,
\ee
\be
N_{m_a}[\bar{\psi^a}\psi^a]= -cm_aN_{m_a}[\cos\sqrt{4\pi }\Phi^a],
\ee
where $N_{m_a}$ denotes normal-ordering with respect to the fermion 
mass $m_a$, $\Phi^a$ is the family of canonical pseudoscalar 
fields. The appearance of the constant
$c=e^{\gamma }/2\pi $ ($\gamma $ is the Euler constant) 
is due to the use of the specific renormalization scheme.
Obviously, physical quantities do not depend on this choice.
 Then, one can derive
the bosonized version of \r{fermla}:
$$
{\cal L}=\frac{1}{2}F_{01}^2 + 
\sum_{a=1}^{N_f}\frac{1}{2}(\partial_\mu\Phi^a)^2 +
\frac{e}{\sqrt{\pi }}F_{01}\left(\sum_{a=1}^{N_f}\Phi^a + \frac{\theta }
{\sqrt{4\pi }}\right)
$$
\be
\la{phila2}
+ \sum_{a=1}^{N_f}cm_a^2N_{m_a}[\cos\sqrt{4\pi }\Phi^a] + const .
\ee
The different vacua are labelled by the angle parametr $\theta $ and
the relevant constant in \r{phila2} should be adjusted to ensure zero vacuum
energy. After integrating out the electric field $F_{01}$, we arrive at the
effective Lagrangian describing the system of interacting pseudoscalar fields 
defined by
$$
{\cal L}_{eff}= \sum_{a=1}^{N_f}\frac{1}{2}(\partial_\mu\Phi^a)^2 - V_{eff}
$$
\be
V_{eff} =
 \frac{e^2}{2\pi }\left(\sum_{a=1}^{N_f}\Phi^a + \frac{\theta }
{\sqrt{4\pi }}\right)^2 
- \sum_{a=1}^{N_f}cm_a^2N_{m_a}[\cos\sqrt{4\pi }\Phi^a]  + const \ .
\label{phila}
\ee
The topological charges of pseudoscalar fields $\Phi^a$ are related to
flavour quantum numbers of fundamental fermions through the relations:
\be
\la{flav}
Q^a=\frac{1}{\sqrt{\pi }}\Phi^a\left |\begin{array}{l}{{}^{+\infty} }\\
{{}_{-\infty} }
\end{array}\right. .
\ee
Classical Euler--Lagrange field equations can be derived,
\be
\partial^2_t \Phi^a - \partial^2_x \Phi^a
+ \frac{e^2}{\pi} \left( \sum_{b=1}^{N_f} \Phi^b +
\frac{\theta}{\sqrt{4\pi}} \right) 
+ c \sqrt{4\pi} m^2_a \sin{\left( \sqrt{4\pi} \Phi^a \right)} = 0 \ .
\label{eula}
\ee

Classical vacua are easy to determine from the requirements 

$\partial V_{eff}/\partial \Phi^a=0$, namely:
\be
\Phi^a=
\frac{1}{\sqrt{4\pi}}
\arcsin{\frac{A}{m_a^2}} + \sqrt{\pi }n^a \ , \ \ \
n^a\in Z \ , \ \ \ \sum_{a=1}^{N_f}n^a=0 \ ,
\ee
where the constant $A$ is subject to the equation
\be
\sum_{a=1}^{N_f}
\arcsin{\frac{A}{m_a^2}} + \theta + c(\frac{2\pi }{e})^2A=0.
\label{cc2}
\ee
All finite-energy (localized) solutions should approach asymptotically the
vacuum:
\be
\Phi^a \stackrel{x \rightarrow \pm \infty}
{\longrightarrow}
\frac{1}{\sqrt{4\pi}} \arcsin{\frac{A}{m_a^2}} + \sqrt{\pi }n^a_{\pm} \ ,
\ee
\be
\partial_{\mu} \Phi^a \stackrel{x \rightarrow \pm \infty}
{\longrightarrow}
0 \ ,
\ee
\be
\sum_{a} n^{a}_{\pm} = 0 \ .
\label{cc1}
\ee
There are no restrictions on our considerations
if we set all $n^a_{-}$ equal zero. Thus, we have 
$n^{a}_{+} = Q^a$, and Eq.(\ref{cc1}) means that all finite-energy
solutions are chargeless (charge screening).  

For $\theta =0$ (no CP breaking) we find
\be
\Phi_{vac}^a=\sqrt{\pi }n^a \ , \ \ \ n^a\in Z \ , \ \ \ 
\sum_{a=1}^{N_f}n^a=0 \ ,
\ee
and we fix the constant in \r{phila} to obtain the potential
\be
\la{weakV}
V_{eff}=\sum_{a=1}^{N_f}cm_a^2N_{m_a}[1-\cos{\sqrt{4\pi }\Phi^a}]+
\frac{e^2}{2\pi }\left(
\sum_{a=1}^{N_f}\Phi^a\right)^2,
\ee

For $\theta \neq 0$, the equation (\ref{cc2}) 
has still a solution corresponding to a classical vacuum.
However, the case of $\theta = \pm \pi$ is special, and the degeneracy
of the vacuum structure appears.

\section{Heavy quarks.}

At first, we consider the case when the Lagrangian (bare) masses
of fundamental fermions are much larger than the scale
of electromagnetic interactions $e$.

Let us divide the Lagrangian (\ref{phila}) into two parts,
\be
{\cal L}^{0}_{eff}= \sum_{a=1}^{N_f}\frac{1}{2}(\partial_\mu\Phi^a)^2 +
 \sum_{a=1}^{N_f}cm_a^2N_{m_a}[\cos\sqrt{4\pi }\Phi^a] \ ,
\label{cc3}
\ee
\be
{\cal L}^{int}_{eff}= 
- \frac{e^2}{2\pi }\left(\sum_{a=1}^{N_f}\Phi^a + \frac{\theta }
{\sqrt{4\pi }}\right)^2 \ .
\label{cc4}
\ee
The first part defines the system of $N_{f}$ sine--Gordon fields.
The weak interactions between them (\ref{cc4}) are important
when fields are close to their vacuum values. 

Since the theory is superrenormalizable, in order to subtract
all ultraviolet divergencies it is enough to replace
the unordered functions of fields in the Lagrangian
by their normal-ordered counterparts. 
Moreover, we do not need to take much care of the renormalization
of quadratic forms of fields (kinetic terms and Coulomb interactions),
because it effects only in the addition of an infinite constant
to the Lagrangian. Thus, everything we want to know about the
ultraviolet renormalization is contained in the prominent
formula derived by Coleman \cite{coleman3}:
\be
\cos{(\beta \Phi)} =
\left( \frac{m}{\Lambda} \right)^{\beta^2 /4 \pi}
N_{m} \left[ \cos{(\beta \Phi)} \right] \ ,
\label{cc5}
\ee
where $\Lambda$ is the ultraviolet cutoff. 
The above formula allows us to compare two
different renormalization scales,
\be
N_{m} \left[ \cos{(\beta \Phi)} \right] =
\left( \frac{\mu}{m} \right)^{\beta^2 /4 \pi}
N_{\mu} \left[ \cos{(\beta \Phi)} \right] \ .
\ee
Thus, the effect of finite renormalization is a multiplication
by the power of mass ratios.  For free fermions (\ref{cc3}),
the anomalous dimension $\beta^2/4\pi$ is equal one, thus 
the operator (\ref{cc5}) acts just like a free fermion mass
operator.

If we want to have the standard form of sine--Gordon Lagrangian,
we need to renormal-order  ${\cal L}^{0}_{eff}$ (\ref{cc3}),
\be
{\cal L}^{0}_{eff} =
 \sum_{a=1}^{N_f} N_{M_a} \left[ \frac{1}{2}(\partial_\mu\Phi^a)^2 +
  \frac{M_a^2}{\beta^2} \cos{(\beta \Phi^a)} \right] \ ,
\label{cc33}
\ee
where $\beta=\sqrt{4\pi}$ and the renormalized mass is:
\be
M_a = \beta^2 c m_a = 2 e^{\gamma} m_a \ .
\ee

As the system of classical field equations derived from
${\cal L}^{0}_{eff}+{\cal L}^{int}_{eff}$
(\ref{cc33}, \ref{cc4}) admits localized
(particle-like) solutions, the semiclassical quantization
can be used to find, at least the lowest, bound states.

There exist two different types of particle-like solutions
for the sine--Gordon equation. The first one refers to solutions
which are time--independent in their rest frames (static, solitonic),
called usually solitons ($Q^a=+1$) and antisolitons ($Q^a=-1$),
\be
\Phi^a = Q^a \frac{4}{\beta} 
\arctan{\left( e^{-M^a (x-x_a)} \right)} \ .
\label{soliton}
\ee
The energy density for such a solution is localized around
some point $x_a$.
The classical masses are ${\cal M}^a_{cl} = 8 M^a/\beta^2$. 
As the  quantum (finite)
corrections do not introduce any new mass scale,
we can denote the quantum mass of solitary solitons as
${\cal M}^a_{qu} = 8 M^a/\beta'^2$, and quantum effects are
reduced to the change of the coupling constant.
 The semiclassical 
WKB quantization around solitons or antisolitons gives the following
effective coupling constant \cite{dashen,rajaraman},
\be
\beta'^2 = \frac{\beta^2}{1 - \frac{\beta^2}{8\pi}}  
\stackrel{\beta = \sqrt{4\pi}} 
\longrightarrow
8 \pi \ .
\label{effb}
\ee
For $\beta=\sqrt{4\pi}$,
the same result has been obtained by the calculation based
on Feynman integrals \cite{ni} 
and it is very close to the numerical value
obtained with variational methods \cite{torres}. Presumably,
for $\beta=\sqrt{4\pi}$ the result of WKB approximation 
can be expected to be exact.

The second type of particle-like
solutions for the sine--Gordon equation
corresponds to periodic in time solutions, called 
breather modes:
\be
\label{breather2}
\Phi^a = \frac{4}{\beta} \arctan{\left(
\frac{\gamma }{\omega }\frac{\sin{\omega t}}{
\cosh{\gamma x}} \right)} \ ,
\ee
where $T=2\pi /\omega $ is a period and 
$\gamma = \sqrt{(M^a)^2 - \omega^2}$.
These solutions are localized and 'topologically chargeless' $Q^a=0$.
Their quantization via WKB methods \cite{dashen,rajaraman}
is analagous to the quantization of the Bohr orbits
of the hydrogen atom. The Bohr--Sommerfeld quantization condition is here:
\be
S + E T = 2 \pi n \ ,
\label{bohrsomm}
\ee
where $n$ is positive integer, $S$ is the action per one period,
and $E$ is the energy. Again, it can be checked that all
quantum effects can be acknowledged in the change of coupling
constant according to (\ref{effb}). The masses of quantum states
produced on the basis of classical breather modes are:
\be
{\cal M}^a_n = \frac{16 M^a}{\beta'^2} 
\sin{\frac{n \beta'^2}{16}} \ ,
\label{breather}
\ee
where $n=1,2,3,...<8\pi/\beta'^2$. For $\beta=\sqrt{4\pi}$ there are
no such states.

After this brief description of the spectrum of the quantized
sine--Gordon system, we can pass to the discussion of the 
quantum states for the Schwinger model (\ref{cc3},\ref{cc4}).
The main result of interactions (\ref{cc4}) is to impose the
zero total charge condition (\ref{cc1}). The lightest bound
states is composed of a pair of a soliton and an antisoliton. Its
mass (up to $e/m_a$ corrections) is given by
\be
{\cal M} = {\cal M}^a_{qu} + {\cal M}_{qu}^b = 
\frac{2 e^{\gamma}}{\pi} (m_a + m_b) \ .
\ee
Let us specify $\theta=0$ and $M^a=M^b=M$, and $Q^a=-Q^b=1$.
Then, the corresponding pair of wave packets (\ref{soliton})
is an exact solution of equations of motion provided
that $x_a=x_b$, i.e. the soliton and the antisoliton are localized
at the same point. It can be immediately checked that the energy
of Coulomb interactions (\ref{cc4}) is zero, so that both 
particle-like waves do not interact together. The state is composed
of a free soliton and a free antisoliton, living together around
the same point $x_a=x_b$.
 But if we try to separate them
for a small distance  $\Delta x = x_a - x_b$, they start to interact
and the energy increase due to
 Coulomb interactions (\ref{cc4}) can be easily
calculated,
\be
\Delta E = \frac{e^2 M}{\pi^2} (\Delta x)^2 \ .
\ee
We have observed that the asymptotic freedom and the confinement 
in the Schwinger model can be already seen at the classical level,
provided that the bosonization has been performed. In this way, the 
bosonization provides us with some kind of the dual description of
the model.

The meson state described above were composed of a pair of
wave packets (\ref{soliton}) corresponding to two particle-like
solutions of different flavours. There exist also states
where 'quark' and 'antiquark' wave packets are of the same
flavour. However, these solutions are of different type, being
breather modes for the system defined by the equations (\ref{eula}). 
Their masses (close to $2 {\cal M}^a$) and their shapes can be found
using the approximate method described in the Appendix.

\section{Light quarks}

In the strong coupling limit, where $e \gg m_a$ it is
much more convenient to perform appropriate change of field
variables:
\be
\label{rotation}
\chi^a=O^a_b\Phi^b + \frac{\theta }{\sqrt{4\pi N_f}}\delta^a_1
\ee
using the orthogonal matrix $O$ defined below:

\bigskip
\begin{eqnarray}
O^1_a & = &\frac{1}{\sqrt{N_f}} (1, 1, \ldots , 1) \nonumber \\
O^2_a & = & \frac{1}{\sqrt{N_f(N_f-1)}} (1, 1, \ldots 1, -N_f+1) 
\nonumber \\
O^3_a & = & \frac{1}{\sqrt{(N_f-1)(N_f-2)}} (1, 1, \ldots 1, -N_f+2,0) 
\nonumber \\
\ldots & & \nonumber \\
O^{N_f}_a & = & \frac{1}{\sqrt{2}} (1, -1, 0, \ldots, 0) 
\label{obrot}
\end{eqnarray}
\bigskip
Using new fields the Lagrangian density (\ref{phila}) takes the form:
\be
\la{chila}
{\cal L}_{eff}=\frac{1}{2}\sum_{a=1}^{N_f}(\partial_\mu \chi^a)^2
-\frac{1}{2}\mu^2
(\chi^1)^2+\sum_{a=1}^{N_f}cm_a^2N_{m_a}[\cos{\sqrt{4\pi }(O^T\chi )^a
-\frac{\theta}{N_f}}] + const ,
\ee
where $\mu^2=N_fe^2/\pi $. The topological charges are defined now
with respect to  $\chi^a$ fields, and they can be
interpreted as the electromagnetic 
charge 
\be
Q=\frac{1}{\sqrt{2\pi }}\chi^1\left |\begin{array}{l}{{}^{+\infty} }\\
{{}_{-\infty} }
\end{array}\right. ,
\ee
and  $SU(N_f)$ hypercharges for other fields. For instance,
in the case of two flavours $N_f=2$
\be
I^3=\frac{1}{\sqrt{2\pi }}\chi^2\left |\begin{array}{l}{{}^{+\infty} }\\
{{}_{-\infty} }
\end{array}\right. 
\ee 
is a third component of isospin. For $N_f=3$ we have two conserved
hypercharges,
\be
Y=\frac{2}{\sqrt{3}}\frac{1}{\sqrt{2\pi }}\chi^2\left 
|\begin{array}{l}{{}^{+\infty} }\\
{{}_{-\infty} } \ ,
\end{array}\right. 
\ee 
\be
I^3=\frac{1}{\sqrt{2\pi }}\chi^3\left |\begin{array}{l}{{}^{+\infty} }\\
{{}_{-\infty} }
\end{array}\right. \ .
\ee 
Following Coleman \cite{coleman}, we can decouple the only heavy field
$\chi^1$. It is straightforward to perform necessary renormal-ordering,
\be
N_{m_a}\left[\cos{\sqrt{\frac{4\pi }{N_f}}(\chi^1-
\frac{\theta }{\sqrt{4\pi N_f}})}\right]=
\left(\frac{\mu }{m_a}\right)^\frac{1}{N_f}N_\mu
\left[\cos{\sqrt{\frac{4\pi }{N_f}}
(\chi^1-\frac{\theta }{\sqrt{4\pi N_f}})}\right]
\ee
and remove $\chi^1$ from \r{chila},
$$
{\cal L}^{light}_{eff}=\frac{1}{2}\sum_{a=2}^{N_f}(\partial_\mu \chi^a)^2
$$
\be
\la{chila2}
+\sum_{a=1}^{N_f}cm_a^2\left(\frac{\mu }{m_a}\right)^\frac{1}{N_f}
N_{m_a}\left[\cos{(-\frac{\theta }{N_f}+
\sqrt{4\pi }\sum_{b=2}^{N_f}O^b_a\chi^b)}\right] + const ,
\ee
The simplest case $N_f=2$ was already elaborated by Coleman \cite{coleman}. 
 He noticed that the Lagrangian density \r{chila2}
is equivalent (for $N_f=2$) to the sine-Gordon theory,
\be
\la{nf2}
{\cal L}^{light}_{N_f=2}=\frac{1}{2}(\partial_\mu \chi^2)^2
+\frac{1}{2\pi }M^2
N_{M}\left[\cos{(\sqrt{2\pi }\chi^2)}\right] ,
\label{qwerty}
\ee
where
\be
M=\left(e^\gamma\mu^{1/2}\sqrt{m_1^2+m_2^2+2m_1m_2\cos\theta }\right)^{2/3}.
\label{masa}
\ee
The quantum states of the sine--Gordon theory were desribed
briefly in the previous section. Here we have $\beta=\sqrt{2\pi}$,
what implies that there are two breathers (\ref{breather2}) with
masses ${\cal M}$ and ${\cal M} \sqrt{3}$ (${\cal M}$ is here
the quantum mass of the soliton). Coleman identified three solutions:
the soliton ($Q=0$, $I_3=+1$), 
the antisoliton ($Q=0$, $I_3=-1$)
and the lighter breather ($Q=0$, $I_3=0$) as the components
of the isotriplet. These are the lightest physical states.
Coleman noticed that they form the degenerate isotriplet even
if the $SU(2)$ flavour symmetry is apparently broken (i.e.
$m_1/m_2$ is very large or very small). However, it is not true that
the $SU(2)$ symmetry is restored here exactly. The differences
in the masses of the isotriplet states arise as we take into 
account the corrections of the order $m_a/\mu$. Note that 
the origin of $I_3=0$ state (the analogue of $\pi^0$) is 
different that the origin of $I_3=\pm 1$
states (the analogues of $\pi^{\pm}$).
Another interesting thing is that the dependence on 
the vacuum parameter $\theta$
is only via the sine--Gordon mass value. As far as the heavier
breather is concerned (that of mass ${\cal M}\sqrt{3}$), Coleman
wrongly \cite{harada}
identified that as the isosinglet state (the analogue
of $\eta$ particle). In fact, the isosinglet state should be 
matched with the non-trivial configuration for $\chi_1$,
so that its mass is of the order $\mu$. Then, the heavier breather mode
should be rather interpreted as a bound state of pions 
(or  $\pi^0$ excitation).

Now, let us consider $N_f=3$ case. We restrict ourselves
to the case when the vacuum angle $\theta $ is zero
($C, P$ symmetries are not broken).
Our effective Lagrangian contains now two light fields,
$$
{\cal L}_{N_f=3}^{light}=\frac{1}{2}(\partial_\mu\chi^2)^2 +
\frac{1}{2}(\partial_\mu\chi^3)^2 - V_{eff} \ , 
$$
$$
V_{eff} = c\mu^{1/3}\left( 
m_1^{5/3}N_{m_1}\left[1- \cos{(\sqrt{2\pi }\chi^3 
+\sqrt{\frac{2\pi }{3}}\chi^2)}\right]\right. +
$$
\be 
m_2^{5/3}N_{m_2}\left[1- \cos{(\sqrt{2\pi }\chi^3 
-\sqrt{\frac{2\pi }{3}}\chi^2)}\right] 
\left.+m_3^{5/3}N_{m_3}\left[1- \cos{\sqrt{\frac{8\pi }{3}}
\chi^2}\right]\right) .
\la{nf3}
\ee
It gives the following field equations of motion ($a=2,3$),
\be
\partial^2_t \chi^a - \partial^2_x \chi^a
+ \frac{\partial V_{eff}}{\partial \chi^a} = 0 \ .
\label{eula2}
\ee
At first, we discuss the case when all fermion masses are equal,
namely $m_1=m_2=m_3=m$, so that $SU(3)$ flavour symmetry remains
unbroken.  We list several exact classical solutions, which
correspond to the lightest bound states. At the begining, note
that if the field $\chi^2$ takes its vacuum value, say $\chi^2=0$,
the  equations (\ref{eula2}) reduce to the sine--Gordon equation
for the field $\chi \equiv \chi^3$,
\be
\partial^2_t \chi - \partial^2_x \chi
+ \frac{M^2}{\sqrt{2\pi}} \sin{(\sqrt{2\pi} \chi)} = 0 \ ,
\label{chi}
\ee
where 
\be
M = \left(2 e^{\gamma} \mu^{1/3} m^{5/3} \right)^{1/2} \ .
\label{masa_}
\ee
Therefore, a soliton, an antisoliton and the lighter one of breathers represent
here three solutions with equal masses, and quantum numbers
$Y=0$ and $I_3=+1,-1,0$ respectively ('pions').  Because of the exact $SU(3)$
flavour symmetry, there are more solutions of the same mass as pions.
In order to obtain them, we need to put $\chi^2/\sqrt{3} \pm \chi^3 =0 $
and observe that $\chi \equiv \chi^2/\sqrt{3} \mp \chi^3 $ 
satisfies the equation (\ref{chi}) (the sign corresponds to
two alternatives). Solitons an antisolitons constructed in these two cases
describe four bound states with quantum numbers $Y=\pm 1$ and
$I_3 = \pm 1/2$ ('kaons'). On the other hand, periodic solutions (breathers)
for both alternative cases can be identified, so that the lighter breather
gives us the eighth state of the pion mass ('$\eta_8$-particle').
 In this way, we have completed 
the whole meson octet, being the family of the lightest physical
particles. Let us remark that this symmetric case allows to
construct further exact solutions. Heavier breathers give next
three particles (mass/octet mass$=\sqrt{3}$, $Y=0,\pm 1$, $I = 0$).
Another example we obtain if we assume $\chi^3=0$ and construct
the soliton solution for the corresponding equation for $\chi^2$
(mass/octet mass $=3/2 + \sqrt{3}\log{(5+\sqrt{24})}$, $Y=\pm 2$
and $I =0$).

We now turn to the discussion of the case when
$ m\equiv m_1=m_2$ and $\mu\gg m_3 \gg m $, i.e. the 'strange'
quark  is much heavier but still below the scale of
interactions. The states of the lowest isotriplet satisfy the equation
\r{chi} with the same assignment of the solutions as before.
The lowest isodoublet states can be constructed in the similar way
as in the case of equal masses. The only difference is that 
the combinations of the fields $\chi^2/\sqrt{3}\pm\chi^3$ do not take
vacuum values, but they  remain still small with respect to
$\chi \equiv \chi^2/\sqrt{3} \mp \chi^3 $. Solving the static equations
of motion \r{eula2} (using methods given in the Appendix) for these variables
we found that the mass ratio between isotriplet states (pions) and 
isodublet states (kaons) is approximately 
$\pi^2\sqrt{r}/4\sqrt{3}$ where
$r=(m_3/m)^{5/3}$.

Finally, we discuss the case when
$ m\equiv m_1=m_2$ and $m_3 \gg \mu \gg m $, i.e. 
one of the flavours refers to heavy quarks. 
In this case we rotate both of the light fields $\Phi^a$ ($a=1,2$) through the 
transformation \r{rotation} into the new field variables $\chi^a$, and
heavy field $\Phi^3$ remains here as the third field variable.
 Using the new fields, our effective lagrangian \r{phila} 
(with $N_f=3$ and $\theta =0$) takes the form:
$$
{\cal L}_{eff}=\frac{1}{2}(\partial_\mu \chi^1)^2+
\frac{1}{2}(\partial_\mu \chi^2)^2+\frac{1}{2}(\partial_\mu \Phi^3)^2-V_{eff}
$$
$$
V_{eff}=\frac{e^2}{\pi }\left(\chi^1+\frac{\Phi^3}{\sqrt{2}}\right)^2-
cm^2N_m[\cos{\sqrt{2\pi }(\chi^1+\chi^2)}+
$$
\be
\cos{\sqrt{2\pi }(\chi^1-\chi^2)}]-
cm_3^2M_{m_3}\cos{\sqrt{4\pi }\Phi^3}+const.
\ee
To consider the lowest states, let us put the heavy field in its vacuum state
$\Phi^3=0$. Then, we derive from the equations of motion 
that field $\chi^1$ 
is in its vacuum state as well,
and field $\chi^2$  satisfies the sine--Gordon equation
with the mass $M=\sqrt{4\pi c} m$ and $\beta=\sqrt{2\pi}$.
 The three solutions of the equal mass to this equation, a soliton, an
antisoliton and one of the breathers represent the three 
partners of the isotriplet  (as it was described after
Eq.\r{chi}). The other light states with a nontrivial contribution from the 
heavy
field can also be constructed. These are isodublets $I^3=\pm1/2$ with nonzero
flavour number  (7) $Q^3=\pm 1$, built on the static solutions of the equation 
of motion. Since the flavour symmetry is broken,
the mass of this state differs from that of the isotriplet. We found 
the approximate value of the 
mass ratio isotriplet mass/isodoublet mass = $\pi^2m_3/(12m)$.
 We also calculated the mass of the $\eta$-particle
 (the appropriate solution is based on the breather), and
 its value calculated up to the first order
 of the method described in the Appendix, is consistent
 with the value predicted by Gell-Mann--Okubo mass formula
$m_\pi^2+3m_\eta^2=4m_K^2$ with the accuracy of 10 per cent.

\section{Summary}

In this paper, we have described the lightest physical
meson states formed from the fundamental fermion fields
in the Schwinger model. Of course, the picture is dependent of
the hierarchy between the quark masses and the scale of
interactions. For heavy quarks, we have noticed that
both confinement and asymptotical freedom can be anticipated
at the classical level. Physical quantities are analytical
both in the coupling constant and in the inverses of quark
masses, so that both parameters can be used to define
perturbative expansions. To describe the physics of light
quarks governed by the Schwinger model, we find it convenient
 to use
a different set of bosonic field variables. 
If all light quark masses are equal (i.e. $SU(N)$
flavour symmetry occurs), then we have checked for
$SU(2)$ and $SU(3)$ examples that the lightest physical
states belong to the mesonic $SU(N)$ multiplets, excluding always
the singlet state. The mass of the singlet state 
lies near the scale of interactions (i.e. the Schwinger mass). 
It is important  to note here  that
the members of the same multiplet correspond
usually to several different types of
classical solutions of the bosonized model.
 In the case of $SU(2)$
symmetry, the breaking of flavour
symmetry can be noticed only as a 'hyperfine' splitting
of the lightest multiplet ('hyperfine' means here being of the
order of the inverse of the Schwinger mass).
 In the case of $SU(3)$ symmetry
(and  higher groups), the situation is different.
When the flavour symmetry is broken, we notice 
immediately within the lightest multiplet the 
splitting of states corresponding to different isospins.

{\Large \bf The Appendix}

We describe here the general approximate method to derive particle-like
solutions for the given system of non-linear field equations. We assume
that we are dealing with two types of particle-like solutions:
static solutions (solitons) and  periodic solutions (breathers).
We present the method looking at the simple example of the exactly solvable
sine--Gordon equation, but there no obstacles if put it into use
in more complicated cases.

The sine--Gordon equation reads,
\be
\partial^2_t \Phi - \partial^2_x \Phi
+ \frac{M^2}{\beta} \sin{(\beta \Phi)} = 0 \ .
\label{sinegord}
\ee
 First, we are attempting to find a static solution localized
around some point $x_0$. We specify the asymptotic conditions
$\Phi(x=-\infty)=0$ and $\Phi(x=+\infty)=2\pi/\beta$. Far away from
the localization point $x_0$ the field $\Phi$ is close to its vacuum
value, and the equation  (\ref{sinegord}) can be linearized.
We write down the linear differential equations for $x << x_0$
and for $x >> x_0$ respectively. Having imposed proper asymptotic
conditions, we glue both solutions together at the point $x_0$.
The final result yields,
\be
\Phi = \left\{ \begin{array}{ll}
\frac{\pi}{\beta} e^{M (x-x_0)} & \mbox{if $x \leq x_0$} \\
\frac{2\pi}{\beta} - \frac{\pi}{\beta} e^{-M (x-x_0)} 
& \mbox{if $x \geq x_0$} 
\end{array} \right.
\ee
It gives the mass $\pi^2 M/\beta^2$, being close to the exact
result $8 M/\beta^2$.
 Using this procedure, one can calculate further
corrections.

To find periodic solutions (breather modes) we need to use
some more tricky procedure. Any periodic solution can be
expanded in the Fourier series,
\be
\Phi = \sum_{n=1}^{\infty} \phi_n(x) \sin{(\omega n t)} \ ,
\label{fourier}
\ee
where $\omega = 2\pi/T$. We restrict here to solutions
antisymmetric in time, $\Phi(-t,x)=-\Phi(t,x)$.
It allows us also to write down the classical energy 
in the following way:
\be
E = \frac{1}{2} \int_{-\infty}^{+\infty} \, dx \ 
\left[ \partial_t \Phi 
\Big|_{t=0} \right]^2 \ .
\label{energybr}
\ee
The Bohr--Sommerfeld quantization condition (\ref{bohrsomm})
reads,
\be
\int_{-\infty}^{+\infty} \, dx \ \int_{0}^{T} \, dt \
\left[ \partial_t \Phi \right]^2 =
2 \pi N \ .
\label{bs2}
\ee
The quantum effects will come to drive the coupling constant 
(\ref{effb}). To find a classical solution being the starting
point for the WKB quantization procedure \cite{dashen},
as a first approximation we take only the first term
in the Fourier expansion (\ref{fourier}). The lowest
frequency $\omega$ is derived from (\ref{bs2}) (for $N=1$),
\be
\omega 
\int_{-\infty}^{+\infty} \, dx \phi_1^2(x) = 2 \ ,
\ee
where the Fourier coefficient $\phi_1$ is calculated from
some ordinary differential equation, which is dependent
of the parameter $\omega$.
The energy (mass) (\ref{energybr})
corresponding to the lowest breather mode  in this
first approximation is given by the following simple formula,
\be
E = \omega \ .
\ee
This is nothing else but the famous de Broglie relation.
Here, this approximate formula is as well as the harmonic
approximation of the breather mode solution is good enough.
If we take the exact solution, we can verify that
$E/\omega = 2 \sqrt{3} /\pi \cong 1.102 $, so that it is
pretty accurate. But it is important to stress
that the above method allows to
calculate further corrections if one wishes.

\end{document}